\documentclass[english,12pt]{article}
          \usepackage{babel}
          \usepackage[T1]{fontenc}
          \usepackage[latin2]{inputenc}
          \usepackage{pslatex}
\begin{document}

\title{\bf Nova V1974 Cygni - results of the 1997 campaign}
\author{Arkadiusz ~O~l~e~c~h}
\date{Nicolaus Copernicus Astronomical Center, \\ 
Polish Academy of Science, \\
ul.~Bartycka~18, 00-716~Warszawa, Poland\\ 
{\tt e-mail: olech@camk.edu.pl}}

\maketitle

\begin{abstract}

This report analyzes the $I$-band CCD photometry of Nova V1974 Cygni
from the 1997 observational season. The analysis shows that both
short-term modulations with periods 0.0813 and 0.085 days are still
present in the light curve of the star. We confirmed the stability of
the shorter period which is interpreted as the orbital period of the
binary system. Its value, determined using the $O-C$ residuals, is 
$P_{orb}=0.08125873(23)$ days = $117.0126(3)$ min. The longer period,
which appeared in the light curve in 1994, was decreasing until the
beginning of 1995 but then started to increase quite rapidly. In
October 1996 the value of the period was $122.67\pm0.02$ min. Until the next
observing run the period significantly decreased. Its value, determined
from our observations performed in July 1997, was $121.87\pm0.12$ min. This
means that the rate of change of the period in 1996-1997 was as high
as $\dot P \approx 10^{-6}$. Such a rapid change of the period requires
a large amount of rotational kinetic energy, if we assume that a 122-min
periodicity is the rotation period of a white dwarf. Thus the more probable
explanation is the hypothesis is that the longer period including a
superhump period is caused by the precession of an accretion disc surrounding
a white dwarf primary.

\noindent {\bf Key words:} Stars: individual: V1974 Cyg -- binaries:
close -- novae, cataclysmic  variables
\end{abstract}

\section{Introduction}

V1974 Cygni (Nova Cygni 1992) was one of the brightest galactic novae of
the 20th century. At the end of February 1992 it reached its maximum
brightness equal to $V=4.3$ mag.

During subsequent years Nova Cygni 1992 was intensively observed using
several techniques. The X-ray flux of the object indicated high luminosity
at the level of $10^{37}$ erg~s$^{-1}$ caused by hot material burning in
thermonuclear reactions at the surface of a white dwarf (Krautter et al.
1996).

The earliest photometric observations obtained by DeYoung and Schmidt
(1993, 1994) revealed the presence of a light curve modulation with a
period of 0.0813 days (117 min) and an amplitude equal to 0.16 mag.

In 1994, Semeniuk et al. (1994) discovered another periodicity in the
light curve of V1974 Cygni with a period equal to 0.085 days (122 min).
Subsequent observations showed that this period is not stable and
changes with $\dot P$ at a level of ~$-10^{-6}$ (Semeniuk et al. 1995,
Olech et al. 1996, Skillman et al. 1997).

The 117 min stable period was interpreted as the orbital period of the
system (DeYoung and Schmidt 1994, Semeniuk et al. 1995, Olech et al.
1996). The nature of the 122 min period still remains unexplained. Until
now, two major explanations have been considered: 1) it might be either
a superhump period of a SU UMa permanent superhumper star (Leibowitz et
al. 1995, Skillman et al. 1997, Retter et al. 1997); or 2) it might be
either the rotation period of the magnetized white dwarf in the binary
system (Semeniuk et al. 1995, Olech et al. 1996).

In this paper we present the results of CCD $I$-band photometry of
V1974 Cygni performed during the end of the observational season 1996
and for the whole season of 1997.

\section{Observations}

The reported observations of V1974 Cygni were obtained on nine nights
during the period from October 4, 1996 to July 15, 1997 at the Ostrowik
station of the Warsaw University Observatory.

The 60-cm Cassegrain telescope, equipped with a Tektronics TK512CB back
illuminated CCD camera, was used. The scale of the camera was 0.76"/pixel
providing a $6.5'\times 6.5'$ field of view. The full description of the
telescope and camera was given by Udalski and Pych (1992).

We have monitored the star in the Cousins $I$ filter. The exposure times
were 180 or 240 seconds depending on the atmospheric conditions, the
seeing and the altitude of the object.

A full journal of our CCD observations of V1974 Cygni is given in Table
1.

\begin{table}[h]
\caption{\sc Journal of the CCD observations of V1974 Cygni}
\vspace{0.1cm}
\begin{center}
\begin{tabular}{|l|c|c|c|}
\hline
\hline 
Date & Time of start & Length of & Number of \\
     & 2450000. + & run [hr] & $I$-band frames \\
\hline
\hline
1996 Oct 04/05 & 361.246 & 6.12 & 89 \\
1997 Jul 04/05 & 634.346 & 2.90 & 31 \\
1997 Jul 06/07 & 636.345 & 1.00 & 14 \\
1997 Jul 09/10 & 639.344 & 4.27 & 57 \\
1997 Jul 10/11 & 640.348 & 4.46 & 42 \\
1997 Jul 11/12 & 641.366 & 4.08 & 59 \\
1997 Jul 12/13 & 642.353 & 4.03 & 54 \\
1997 Jul 13/14 & 643.334 & 3.50 & 23 \\
1997 Jul 15/16 & 645.409 & 3.05 & 45 \\
\hline
Total          &   --   & 33.41 & 414 \\ 
\hline
\hline
\end{tabular}
\end{center}
\end{table}

\section{Data Reduction}

As in our previous papers describing the photometric behavior of V1974
Cygni (Semeniuk et al. 1994, 1995, Olech et al. 1996) the initial
reduction of the CCD frames (subtracting BIAS, DARK and FLAT-FIELD
images) was done using the standard routines in the IRAF CCDPROC
package. \footnote{IRAF is distributed by the National Optical Astronomy
Observatory, which is operated by the Association of Universities for
Research in Astronomy, Inc., under a cooperative agreement with the
National Science Foundation.}

\vspace{18.5cm}

\includegraphics{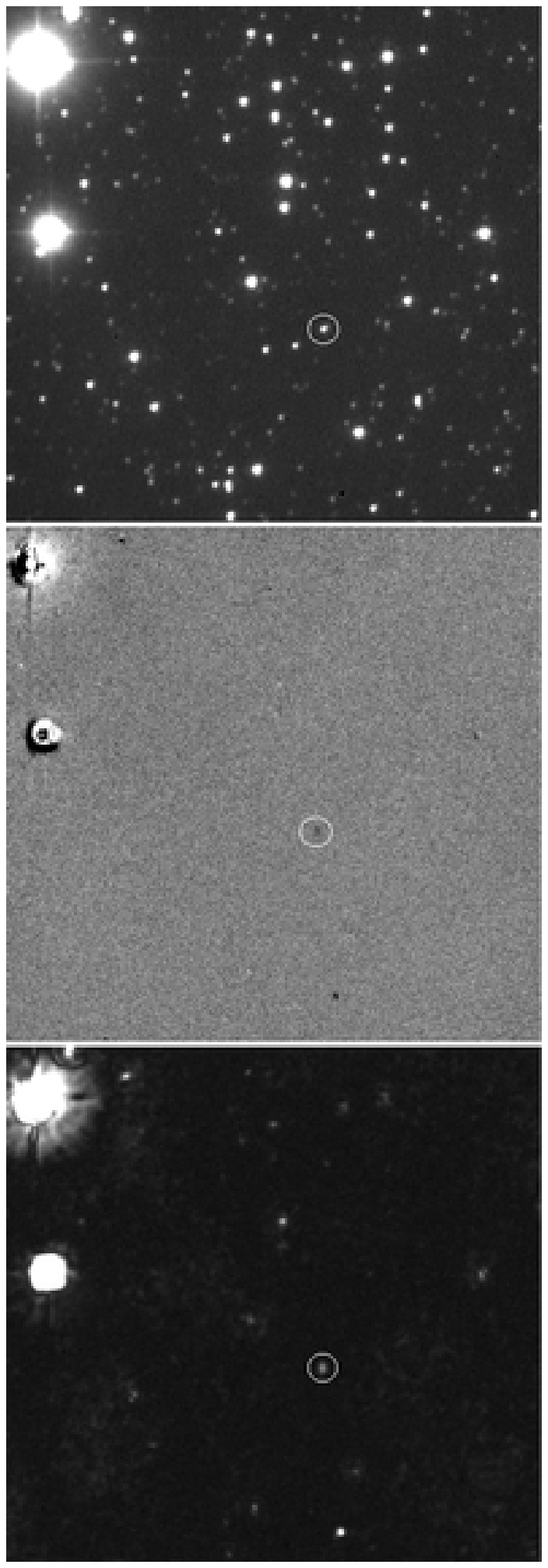}

\begin{figure}[h]
\caption{\sf {\it Upper panel:} The reference image stacked using the 14
best frames. ~{\it Middle panel:} One of the images obtained by subtracting
the convolved reference image. Black dots are the possible variable 
stars. ~{\it Lower panel:} {\tt ABS} image obtained by stacking absolute  
values of all frames produced during the subtraction. The circle matches
the position of V1974 Cygni.}
\end{figure}
\clearpage

A further reduction was done using a newly developed image subtraction
method ISIS (Alard and Lupton 1998, Alard 2000). ISIS is based on the
fast optimal image subtraction algorithm. Subtraction is done between
the reference image (obtained by stacking several best seeing and low
background images) and each image from the data set transformed to the
coordinate grid of the reference image. Before subtraction the reference
image is modified to exactly match the seeing of each image from
the data set. This is done by finding the convolution kernel and
difference in background levels between the reference and subtracted
image. There are more details in the papers of Alard and Lupton (1998)
and Alard (2000).

It was shown by Alard (1999) and Olech et al. (1999) that ISIS works
very well even in very dense regions of the Galactic bulge and in the
cores of the globular clusters. In the first applications ISIS produced
light curves with a quality significantly better than obtained by
traditional methods such as DaoPHOT and DoPHOT.

During recent years the image subtraction method was widely used for
several astronomical objects such as globular clusters (Olech at al.
1999, Kaluzny, Olech and Stanek 2001, Kopacki 2001), open clusters
(Mochejska et al. 2002), galaxies (Mochejska et al. 2001), quasars
(Wo\'zniak et al. 2000) or microlensing events (Alard 1999, Soszy\'nski
et al 2001).

The field of V1974 Cygni located in the Galactic plane is quite rich in
stars. There are a few hundreds stellar objects in the $6.5'\times6.5'$
field of view of the Ostrowik telescope thus CCD frames of the V1974 Cygni
are optimal for reduction using the ISIS package.

In the first step, all frames were transformed to a common $(x,y)$
coordinate grid. Next we constructed the reference image stacking 14 of
the best exposed and low background images. The obtained reference image
is shown in the upper panel of Fig. 1.

Image subtraction was then applied to all the frames from our data set.
For each CCD frame, the reference image was convolved with a kernel to
match its point spread function (PSF), and then it was subtracted from
the frame producing set of difference images. One such images is
shown in the middle panel of Fig. 1.

\vspace{8.5cm}

\includegraphics{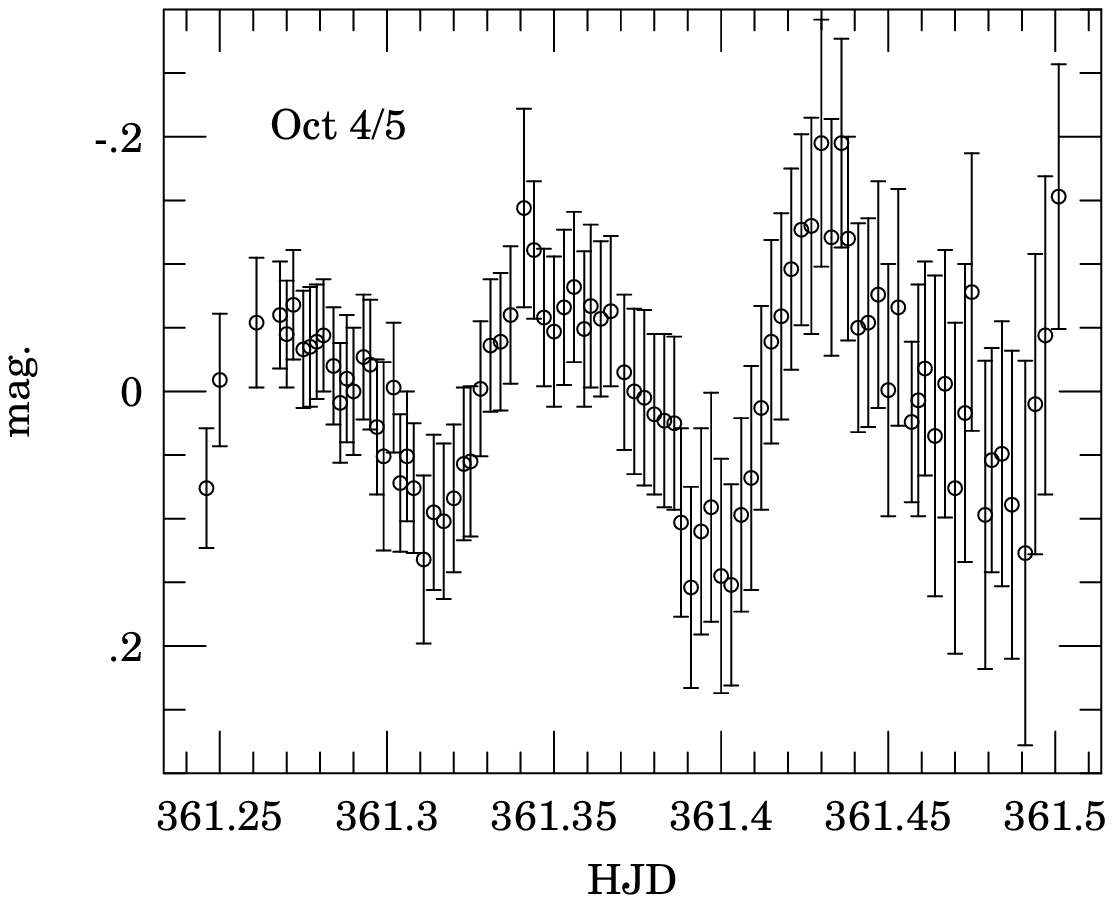}

\begin{figure}[h]
\caption{\sf The light curve of V1974 Cygni observed during the night of
October 4/5, 1996.}
\end{figure}
\clearpage

At this stage the standard ISIS package produces {\tt ABS} image, which
is built by computing a median of absolute deviations on all subtracted
images. Such an image obtained for our data set is shown in the lower
panel of Fig. 1. The detection of the potential variables on such a
frame can be done using for example the {\tt find} algorithm of DaoPHOT
software (Stetson 1987).

\vspace{19.1cm}

\includegraphics{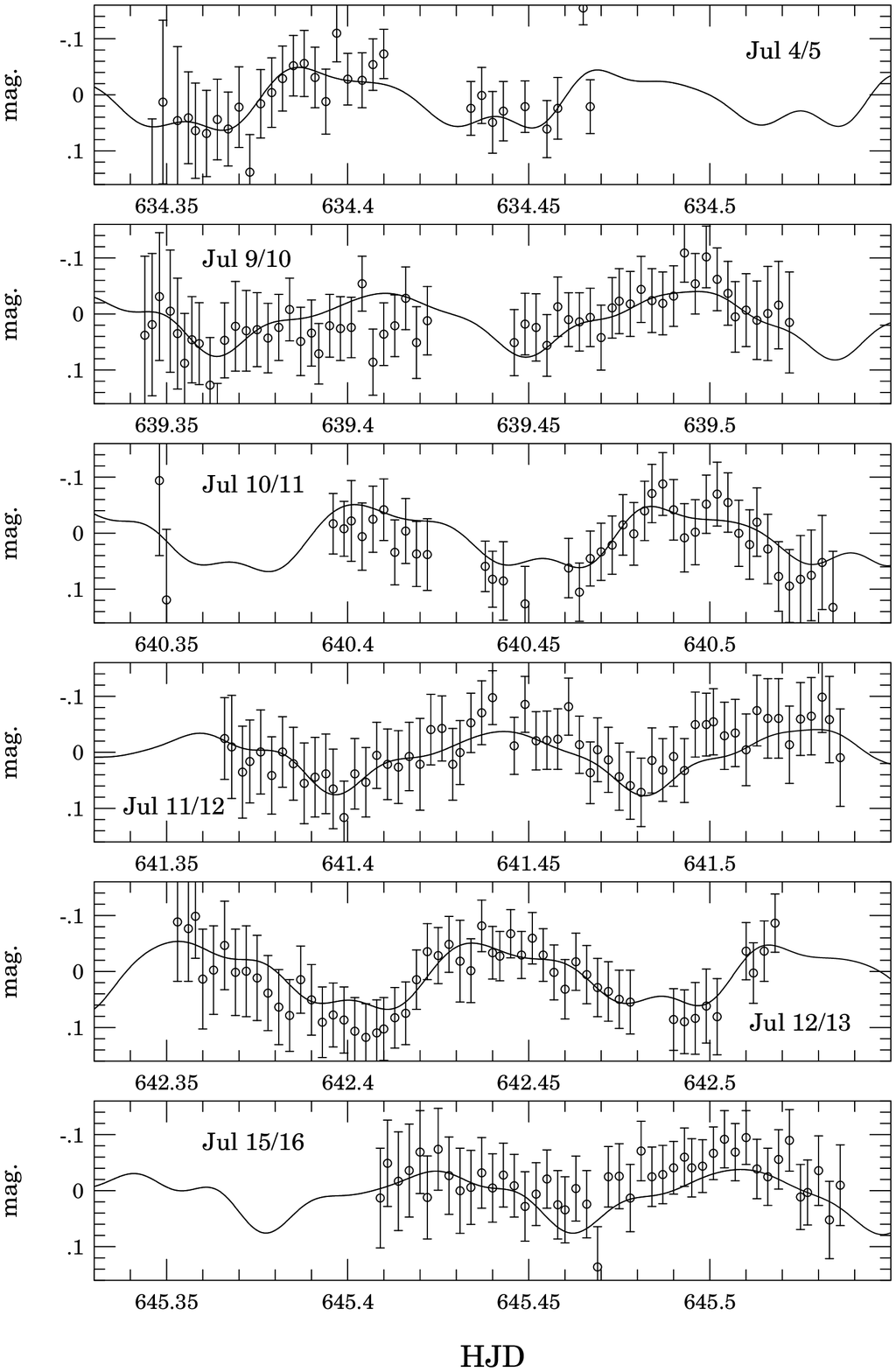}

\begin{figure}[h]
\caption{\sf The light curve of V1974 Cygni observed during six nights
in July 1997. The solid line is the fit given by equation (1).}
\end{figure}
\clearpage

In our case we have already known the position of the variable which we
were interested in. In Fig. 1 the variable V1974 Cyg is thus marked by
an open circle. One can clearly see that it is detectable by both its
difference and {\tt ABS} image.

The main disadvantage of using the ISIS package is obtaining the light
curves in ADU and not in the magnitudes. Thus we should convert the light
curves from ADU to magnitudes. For this purpose, the reference frame was
reduced with the DaoPHOT/ALLSTAR package (Stetson 1987). Knowing the
instrumental magnitude of the variable produced by the ALLSTAR routine and
ADU flux of this object at the reference frame, the ADU light curve can be
transformed into magnitudes.

The resulting light curve from the night of October 4/5, 1996 is
displayed in Fig. 2 and light curves from six of eight July 1997 nights
are shown in Fig. 3.

\section{Power Spectrum Analysis of 1997 Data}

To obtain power spectra of the 1997 photometric observations of V1974
Cygni we have used the AoV (Schwarzenberg-Czerny 1989) statistic. Before
calculating the power spectrum we have removed the nightly mean and a
longer-scale trend from each individual run. Fig. 4 presents results of
the analysis for all July 1997 nights.

Upper panel of Fig. 4 shows the AoV power spectrum for the raw light curve.
The solid ticks in the figure indicate the frequencies 11.807 and 12.302
cycles/day. The corresponding periods are equal to $0.08470\pm0.00014$
days ($122.0\pm0.2$ min) and $0.0813\pm0.0002$ days ($117.1\pm0.3$ min).

Knowing both periodicities we have fitted our light curve with the
Fourier sine series in the following form:

\begin{equation}
mag = A_0 + \sum^{4}_{j=1} A^1_j\cdot\sin({{2\pi j t}\over{P_1}} +
\varphi^1_j) + \sum^{4}_{j=1} A^2_j\cdot\sin({{2\pi j t}\over{P_2}} 
+ \varphi^2_j)
\end{equation}

\noindent Then, knowing $P_2$, $A^2_j$ and $\varphi^2_j$ from the raw
light curve we have removed the variability with a period $P_2$. For
such prewhitened light curve we have computed the AoV spectrum which is
displayed in Fig. 4b. The only frequency left is $f_1=11.807$ cycles/day
corresponding to the period $P_1=0.0847$ days.

On the other hand having $P_1$, $A^1_j$ and $\varphi^1_j$ one can make
prewhitening of the light curve with period $P_1$. The result of such an
operation is shown in Fig. 4c. In this case the only frequency that left
is $f_2=12.302$ cycles/day which corresponds to the period $P_1=0.0813$
days.

Finally we have prewhitened our raw light curve of V1974 Cygni removing
the variability with both periods $P_1$ and $P_2$. The AoV power
spectrum of such prewhitened light curve is shown in Fig. 4d. There are
no clear peaks in this graph indicating that $f_1$ and $f_2$ are
sole frequencies present in the light curve of Nova Cygni 1992.

Knowing the values of both frequencies detected in the light curve of
the star we were able to determine the amplitudes and shapes of light
variations. The upper panel of Fig. 5 shows the raw light curve of V1974
Cygni phased with more distinct period $P_1=0.08470$ days. Due to the
presence of the weaker signal characterized by period $P_2=0.0813$ days
phased light curve shows significant scatter. Thus, based on equation
(1), we removed the variations with $P_2$ and such a prewhitened light
curve again phased with period $P_1$. The result is shown in the middle
panel of Fig. 5. The points in this graph were additionally averaged in
0.025 phase bins. The solid line corresponds to the part of the equation
(1) containing terms with $P_1$. The full amplitude of the modulation is
$0.090\pm0.013$ mag.

\clearpage

~

\vspace{19.5cm}

\includegraphics{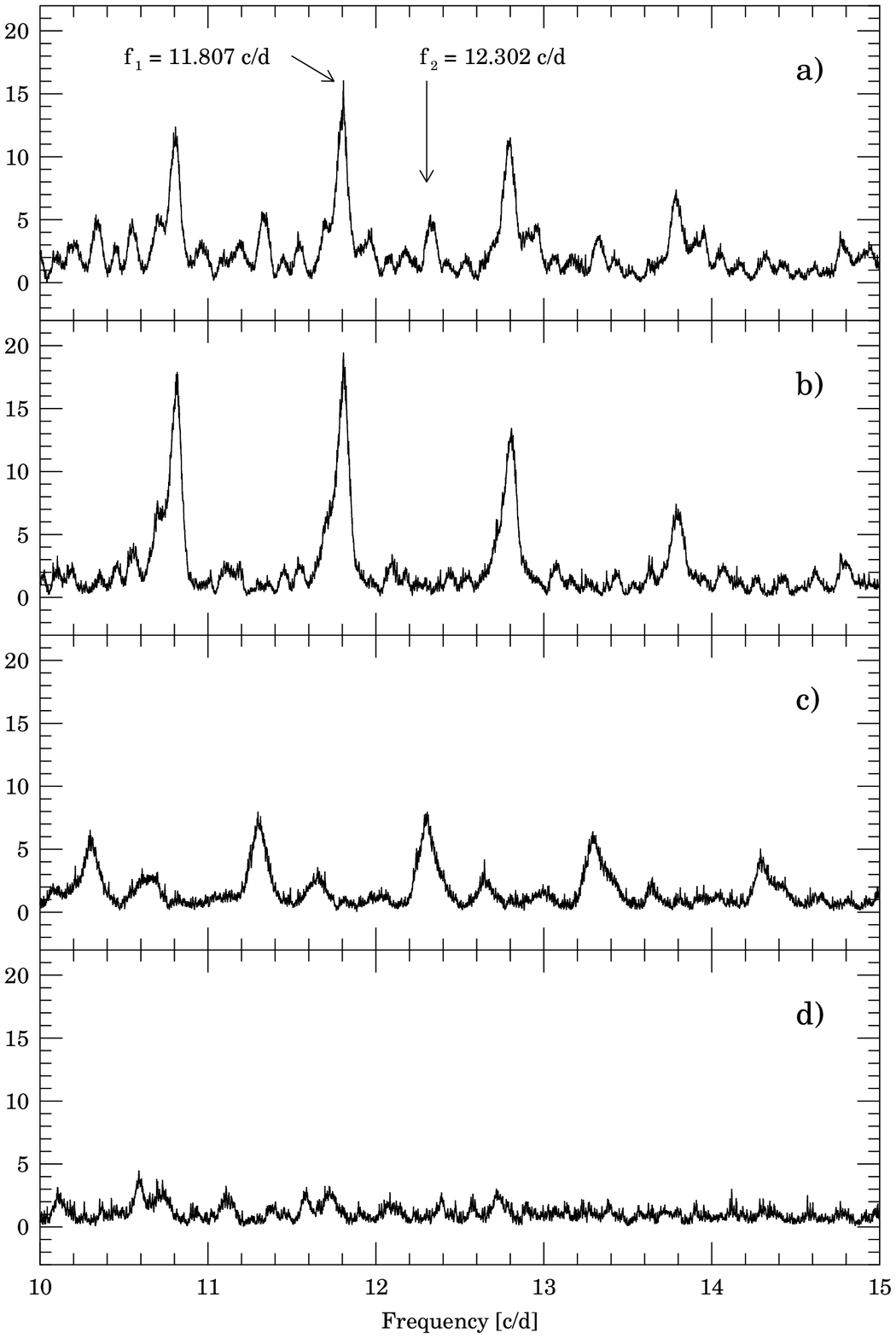}

\begin{figure}[h]
\caption{\sf {\bf a)} AoV power spectrum of the raw light curve of V1974
Cygni from 1997. ~{\bf b)} AoV power spectrum prewhitened with
variability with period $P_2$. ~{\bf c)} AoV power spectrum prewhitened
with  variability with period $P_1$. ~{\bf d)} AoV power spectrum
prewhitened with variability with periods $P_1$ and $P_2$.}
\end{figure}

\clearpage
~

\vspace{19.1cm}

\includegraphics{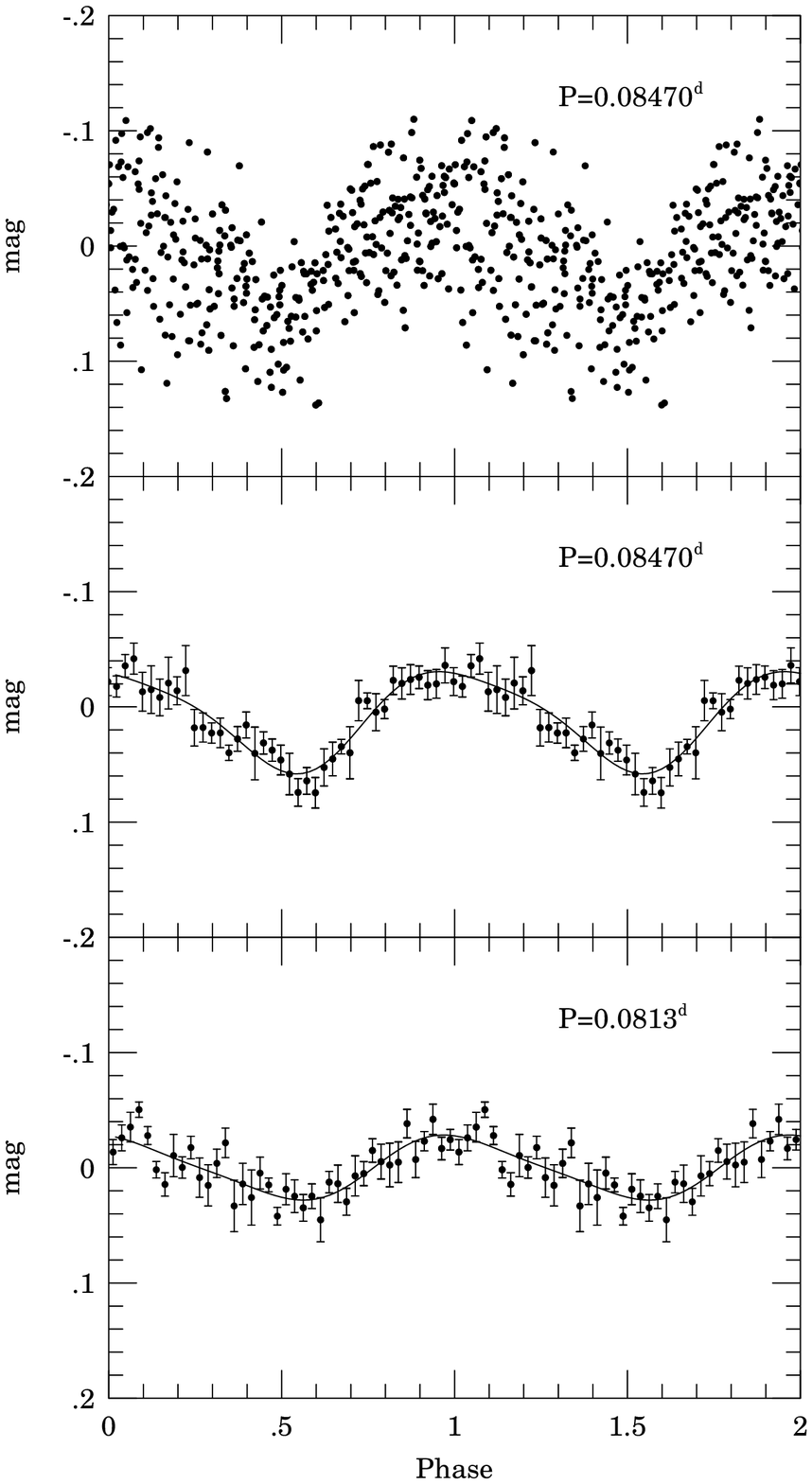}  

\begin{figure}[h]
\caption{\sf The phased light curves of V1974 Cygni from July 1997. {\it
Upper panel:} Raw light curve phased with period 0.08470 days. ~{\it
Middle panel:} Light curve prewhitened with period 0.0813 days and
phased with period 0.08470 days. ~{\it Lower panel:} Light curve
prewhitened with period 0.08470 days and phased with period 0.0813 days.
Solid lines correspond to the proper terms of the fit given by equation (1).}
\end{figure}

\clearpage

The lower panel of Fig. 5 shows a phased light curve with removed
variations with period $P_1$ and its three harmonics. Again the points
in this graph are the average values of the 0.025 phase bins. The full
amplitude of this modulation is $0.057\pm0.015$ mag.

\section{O -- C Diagrams}

\subsection{The 0.0813-day Period}

The $O-C$ analysis of times of extrema in a variable star is a very
useful tool for tracing the changes of the period of the variable. Such
an analysis, performed by DeYoung and Schmidt (1994), Semeniuk et al.
(1995), Olech et al. (1996) and Skillman et al. (1997), showed that the 
0.0813-day period of Nova Cygni 1992 is stable and can be interpreted as
an orbital period of the binary system. Additionally Olech et al. (1996)
pointed out that in 1994 this period underwent a slight change from
0.0812623 to 0.0812580 days, i.e. shortened by about 0.4 seconds. A
possible mechanism of such a shortening of the orbital period may be the
outflow of mass from the system through an outer Lagrangian point.

Now we can derive more definite conclusions using all maxima from the
years 1993-1996 presented in papers by Semeniuk et al. (1995), Olech
et al. (1996), Skillman et al. (1997) and from the years 1996-1997
presented in this paper.

Table 2 contains 69 times of maxima of the 0.0813-day modulation
presented in the above-mentioned papers and in this work. The maximum
from the night of October 4/5 1996 was determined as the average value
of three peaks detected in our data. Because the amplitude of the
modulation (see Fig. 2) was very high, we assumed that during this night
the maxima of both periodicities were occurring at the same time, thus
strengthening the modulations. Light variations detected in the light
curve in July 1997 had a more complicated pattern, thus we decided to
use it for only one maximum determination based on the $P_2$ term of the
fit given by equation (1).

\vspace{6.7cm}

\includegraphics{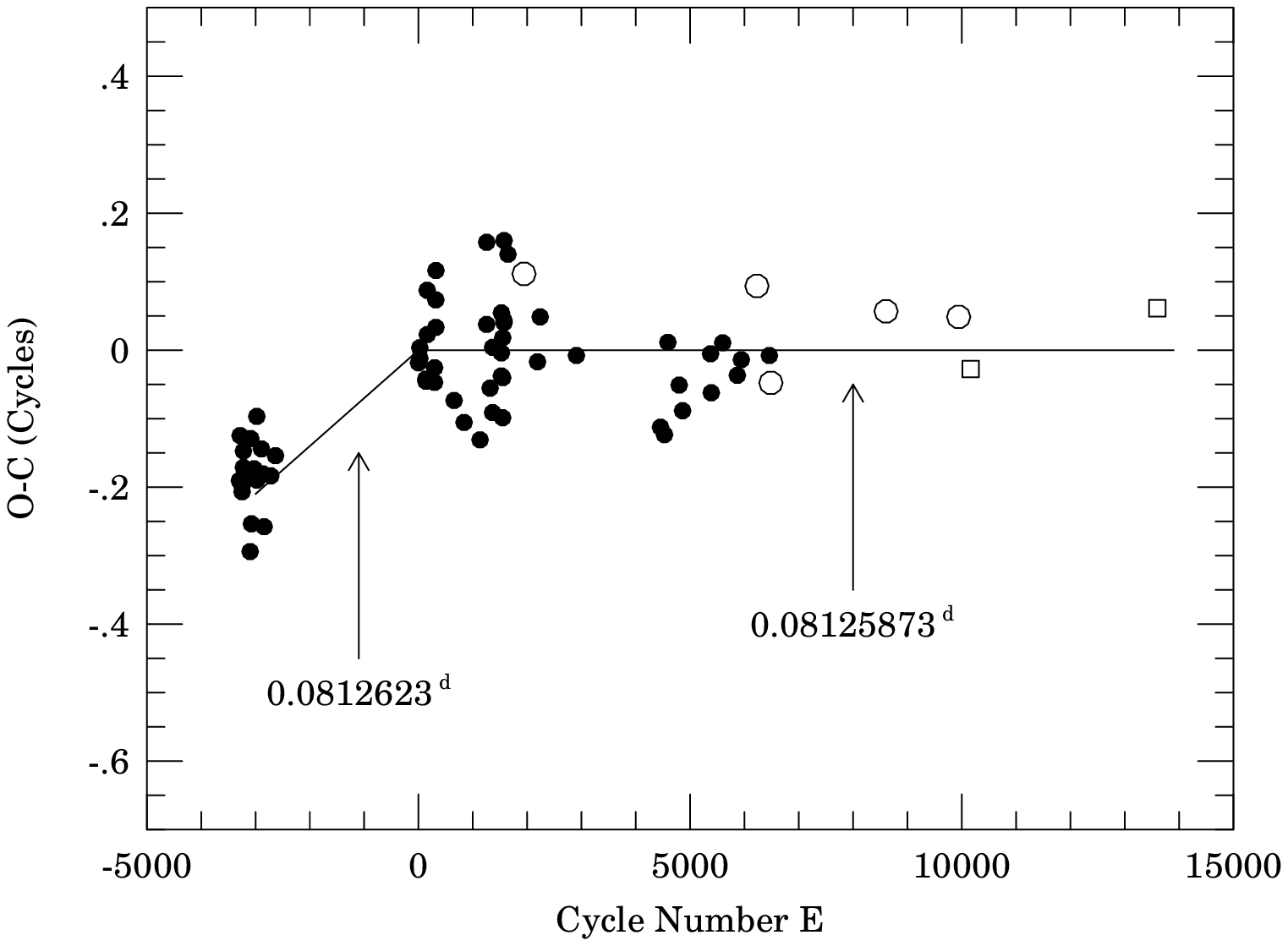}

\begin{figure}[h]
\caption{\sf $O-C$ diagram for times of maxima of the 0.0813-day
(orbital) modulation observed in the years 1993-1997. The $O-C$
residuals were calculated from the ephemeris (2). Black dots represent
data used by Olech et al. (1996), open circles - maxima given by Skillman
et al. (1997) and open squares - maxima determined in this work.}
\end{figure}

\clearpage

\begin{table}[t]
\caption{\sc The 1993-1997 Times of Maxima of the 0.0813-day
Periodicity. 1 - Semeniuk et al. (1995), ~2 - Olech at al. (1996), ~3 - 
Skillman et al. (1997), ~4 - this work.}
\vspace{0.1cm}
\begin{center}
{\small
\begin{tabular}{||c|c|c|c||c|c|c|c||}
\hline
\hline
HJD & E & $O-C$ & Ref. & HJD & E & $O-C$ & Ref. \\
2449000+ & & cycles & & 2449000+ & & cycles & \\
\hline
267.5140 & -3297 & -0.1903 & (1) &    642.534 & 1318 & -0.0554 & (1) \\
268.5757 & -3284 & -0.1246 & (1) &    646.269 & 1364 & -0.0911 & (1) \\
271.4950 & -3248 & -0.1986 & (1) &    646.358 & 1365 & 0.0041 & (1) \\
271.5756 & -3247 & -0.2067 & (1) &    659.356 & 1525 & -0.0376 & (1) \\
273.5287 & -3223 & -0.1712 & (1) &    659.440 & 1526 & -0.0039 & (1) \\
273.6119 & -3222 & -0.1473 & (1) &    659.526 & 1527 & 0.0543 & (1) \\
283.5948 & -3099 & -0.2940 & (1) &     661.220 & 1548 & -0.0986 & (1) \\
284.5833 & -3087 & -0.1292 & (1) &    661.306 & 1549 & -0.0402 & (1) \\
285.5483 & -3075 & -0.2535 & (1) &    661.392 & 1550 & 0.0180 & (1) \\
286.5292 & -3063 & -0.1822 & (1) &    663.263 & 1573 & 0.0432 & (1) \\
289.5365 & -3026 & -0.1733 & (1) &    663.344 & 1574 & 0.0401 & (1) \\
293.5244 & -2977 & -0.0967 & (1) &   663.435 & 1575 & 0.1599 & (1) \\
293.5981 & -2976 & -0.1897 & (1) &    669.284 & 1647 & 0.1399 & (1) \\
300.5088 & -2891 & -0.1441 & (1) &    693.253 & 1942 & 0.1113 & (3) \\
302.5373 & -2866 & -0.1806 & (1) &    713.476 & 2191 & -0.0169 & (1) \\
304.5625 & -2841 & -0.2577 & (1) &    717.463 & 2240 & 0.0485 & (1) \\
314.4821 & -2719 & -0.1835 & (1) &    771.658 & 2907 & -0.0077 & (2) \\
321.554 & -2632 & -0.1541 & (1) &     897.438 & 4455 & -0.1125 & (2) \\
535.438 & 0 & -0.0184 & (1) &        903.369 & 4528 & -0.1234 & (2) \\
537.390 & 24 & 0.0035 & (1) &       908.418 & 4590 & 0.0114 & (2) \\
537.470 & 25 & -0.0119 & (1) &       925.396 & 4799 & -0.0510 & (2) \\
546.406 & 135 & -0.0421 & (1) &       930.431 & 4861 & -0.0884 & (2) \\
546.487 & 136 & -0.0453 & (1) &       972.286 & 5376 & -0.0053 & (2) \\
548.448 & 160 & 0.0874 & (1) &       973.419 & 5390 & -0.0622 & (2) \\
548.524 & 161 & 0.0226 & (1) &       990.408 & 5599 & 0.0107 & (2) \\
559.407 & 295 & -0.0470 & (1) &      1012.344 & 5869 & -0.0367 & (2) \\
559.490 & 296 & -0.0256 & (1) &      1018.359 & 5943 & -0.0139 & (2) \\
561.367 & 319 & 0.0734 & (1) &       1041.689 & 6230 & 0.0936 & (3) \\
561.445 & 320 & 0.0333 & (1) &        1060.289 & 6459 & -0.0078 & (2) \\
561.533 & 321 & 0.1162 & (1) &        1062.561 & 6487 & -0.0477 & (3) \\
588.658 & 655 & -0.0734 & (1) &      1234.838 & 8607 & 0.0567 & (3) \\
603.607 & 839 & -0.1055 & (1) &       1343.399 & 9943 & 0.0485 & (3) \\
627.495 & 1133 & -0.1309 & (1) &      1361.351 & 10164 & -0.0274 & (4) \\
637.341 & 1254 & 0.0375 & (1) &      1640.5632 & 13600 & 0.0611 & (4) \\
637.432 & 1255 & 0.1574 & (1) &       & & & \\
\hline
\hline
\end{tabular}}
\end{center}
\end{table}
\bigskip

\clearpage

The 0.0813-day period behavior in the 1993-1997 time interval is
illustrated in Fig. 6. The $O-C$ residuals displayed in the figure and
shown in Table 2 were calculated with the linear ephemeris:

\begin{equation}
{\rm HJD_{max}} = 2449535.4395 + 0.08125873~E
\end{equation}
\hspace*{213pt} ${\pm}0.0010 \hspace*{5pt} {\pm}0.00000023$ \\
\bigskip

Fig. 6 shows clearly that since the change in 1994 the orbital period
of the binary system is stable and in the years 1994-1997 is equal to
$P_{orb}=0.08125873(23)$ days.

\subsection{The 0.085-day Period}

Contrary to the 0.0813-day period, the longer period showed a more
complicated behavior. Not present in the 1993 data, the 0.085-day period
was the dominant peak in the light curve from the years 1994-1997
(Semeniuk et al. 1994, 1995, Olech et al. 1996, Skillman et al. 1997,
Retter et al. 1997).

It was first demonstrated by Semeniuk et al. (1995) that the 0.085-day
period is changing. Data collected in 1994 clearly showed a decrease of
the period with a rate $\dot P= -7.8\times 10^{-7}$. In the beginning of 1995
the period started to increase (Olech et al. 1996) and was still
increasing with $\dot P= 5\times 10^{-7}$ in 1996 (Skillman et al. 1997).

We decided to use all times of extrema of V1974 Cygni existing in the
literature and determined in this work to trace the behavior of the
0.085-day period in detail. We used maxima given by Semeniuk et
al (1994, 1995), Olech et. al (1996) and Retter et al. (1997). Skillman
et al. (1997) tabulated only times of minima of 0.085-day periodicity
but based on his Fig. 6 we transformed them into maxima adding the 0.45
phase shift to the minima.

Table 3 contains 139 times of maxima of the 0.085-day modulation
presented in above-mentioned papers and determined in this work.

The 0.085-day period behavior in the 1994-1996 time interval is
presented in Fig. 7. The $O-C$ residuals displayed in the figure and
shown in Table 3 were calculated with the linear ephemeris:

\begin{equation}
{\rm HJD_{max}} = 2449535.4320 + 0.085~E
\end{equation}

The fitted curve shown as a solid line corresponds to the ephemeris:

\begin{equation}
{\rm HJD_{max}} = 2449535.4406 + 0.08500527~E - 1.435 \times 10^{-8}~E^2
+ 1.752 \times 10^{-12}~E^3
\end{equation}

Lack of the data between 1995 and 1996 seasons may suggest ambiguity in
cycle count. In our opinion it is not in this case because equation (4)
produces the value of the period equal to 122.66 min which is almost
exactly the same as was reported by Skillman et al. (1997) from 1996
data i.e. $P=122.67\pm0.02$~min. Cycle numerations in 1996 differing by
plus or minus one from the scheme presented in Table 3 give the periods
122.63 and 122.74 min, respectively i.e. marginally consistent with data
of Skillman et al. (1997).

Due to the lack of data in the almost 300-day interval between October
1996 and July 1997 a construction of the $O-C$ diagram for the 1996-1997
data was impossible. 

\clearpage

\begin{table}[h]
\caption{\sc The 1994-1997 Times of Maxima of the 0.085-day
Periodicity. 1 - Semeniuk et al. (1995), ~2 - Olech et al. (1996),
3 - Skillman et al. (1997), 4 - Retter et al. (1997), ~5 - this work.}
\vspace{0.1cm}
\begin{center}
{\tiny
\begin{tabular}{||c|c|c|c||c|c|c|c||}
\hline
\hline
HJD & E & $O-C$ & Ref. & HJD & E & $O-C$ & Ref. \\
2449000+ & & cycles & & 2449000+ & & cycles & \\
\hline
471.540 & -751 & -0.6705 & (1) & 878.34 & 4035 & -0.7882 & (4) \\
535.438 & 0 & 0.0705 & (1) & 878.42 & 4036 & -0.8470 & (4) \\
537.390 & 23 & 0.0352 & (1) & 878.507 & 4037 & -0.8235 & (4) \\
537.470 & 24 & -0.0235 & (1) & 879.34 & 4047 & -1.0235 & (4) \\
546.406 & 129 & 0.1058 & (1) & 879.442 & 4048 & -0.8235 & (4) \\
546.487 & 130 & 0.0588 & (1) & 879.53 & 4049 & -0.7882 & (4) \\
547.424 & 141 & 0.0823 & (1) & 880.380 & 4059 & -0.7882 & (4) \\
548.448 & 153 & 0.1294 & (1) & 880.451 & 4060 & -0.9529 & (4) \\
548.524 & 154 & 0.0235 & (1) & 880.545 & 4061 & -0.8470 & (4) \\
559.407 & 282 & 0.0588 & (1) & 897.449 & 4260 & -0.9764 & (2) \\
559.490 & 283 & 0.0352 & (1) & 903.385 & 4330 & -1.1411 & (2) \\
560.437 & 294 & 0.1764 & (1) & 908.409 & 4389 & -1.0353 & (2) \\
561.367 & 305 & 0.1176 & (1) & 925.319 & 4588 & -1.0941 & (4) \\
561.445 & 306 & 0.0352 & (1) & 925.391 & 4589 & -1.2470 & (2) \\
561.533 & 307 & 0.0705 & (1) & 925.392 & 4589 & -1.2352 & (4) \\
562.395 & 317 & 0.2117 & (1) & 925.484 & 4590 & -1.1529 & (4) \\
562.479 & 318 & 0.2000 & (1) & 925.574 & 4591 & -1.0941 & (4) \\
570.800 & 416 & 0.0941 & (3) & 930.408 & 4648 & -1.2235 & (2) \\
588.658 & 626 & 0.1882 & (1) & 972.301 & 5141 & -1.3647 & (2) \\
603.607 & 802 & 0.0588 & (1) & 973.409 & 5154 & -1.3294 & (2) \\
604.714 & 815 & 0.0823 & (3) & 977.240 & 5199 & -1.2588 & (4) \\
620.270 & 998 & 0.0941 & (4) & 977.318 & 5200 & -1.3411 & (4) \\
620.347 & 999 & 0.0000 & (4) & 977.408 & 5201 & -1.2823 & (4) \\
620.445 & 1000 & 0.1529 & (4) & 977.485 & 5202 & -1.3764 & (4) \\
625.547 & 1060 & 0.1764 & (1) & 990.407 & 5354 & -1.3529 & (2) \\
627.495 & 1083 & 0.0941 & (1) & 993.225 & 5387 & -1.2002 & (4) \\
627.495 & 1083 & 0.0941 & (3) & 1005.28 & 5529 & -1.3764 & (4) \\
637.262 & 1198 & 0.0000 & (1) & 1009.27 & 5576 & -1.4352 & (4) \\
637.341 & 1199 & -0.0710 & (1) & 1010.224 & 5587 & -1.2117 & (4) \\
637.432 & 1200 & 0.0000 & (1) & 1011.47 & 5602 & -1.5529 & (4) \\
642.534 & 1260 & 0.0235 & (1) & 1012.334 & 5612 & -1.3882 & (2) \\
646.269 & 1304 & -0.0352 & (1) & 1012.422 & 5613 & -1.3529 & (4) \\
646.358 & 1305 & 0.0117 & (1) & 1018.375 & 5683 & -1.3176 & (2) \\
650.595 & 1355 & -0.1411 & (3) & 1038.264 & 5917 & -1.3294 & (4) \\
651.295 & 1363 & 0.0941 & (4) & 1038.351 & 5918 & -1.3058 & (4) \\
651.370 & 1364 & -0.0235 & (4) & 1041.251 & 5952 & -1.1882 & (4) \\
651.540 & 1366 & -0.0235 & (3) & 1041.336 & 5953 & -1.1882 & (4) \\
653.662 & 1391 & -0.0588 & (3) & 1060.274 & 6176 & -1.3882 & (2) \\
655.617 & 1414 & -0.0588 & (3) & 1234.939 & 8229 & 0.4941 & (3) \\
659.356 & 1458 & -0.0705 & (1) & 1237.836 & 8263 & 0.5764 & (3) \\
659.440 & 1459 & -0.0823 & (1) & 1239.965 & 8288 & 0.6235 & (3) \\
659.526 & 1460 & -0.0705 & (1) & 1240.822 & 8298 & 0.7058 & (3) \\
661.220 & 1480 & -0.1411 & (1) & 1242.777 & 8321 & 0.7059 & (3) \\
661.306 & 1481 & -0.1294 & (1) & 1243.799 & 8333 & 0.7294 & (3) \\
661.392 & 1482 & -0.1176 & (1) & 1246.531 & 8365 & 0.8705 & (3) \\
663.263 & 1504 & -0.1058 & (1) & 1248.901 & 8393 & 0.7529 & (3) \\
663.344 & 1505 & -0.1529 & (1) & 1250.777 & 8415 & 0.8235 & (3) \\
663.435 & 1506 & -0.0823 & (1) & 1252.910 & 8440 & 0.9176 & (3) \\
665.216 & 1527 & -0.1294 & (4) & 1254.869 & 8463 & 0.9647 & (3) \\
665.298 & 1528 & -0.1647 & (4) & 1256.831 & 8486 & 1.0470 & (3) \\
665.384 & 1529 & -0.1529 & (4) & 1257.849 & 8498 & 1.0235 & (3) \\
669.284 & 1575 & -0.2705 & (1) & 1258.790 & 8509 & 1.0941 & (3) \\
693.170 & 1856 & -0.2588 & (4) & 1258.957 & 8511 & 1.0588 & (3) \\
693.243 & 1857 & -0.4000 & (4) & 1260.747 & 8532 & 1.1176 & (3) \\
693.322 & 1858 & -0.4705 & (4) & 1261.856 & 8545 & 1.1647 & (3) \\
694.273 & 1869 & -0.2823 & (4) & 1262.791 & 8556 & 1.1647 & (3) \\
694.359 & 1870 & -0.2705 & (1) & 1285.608 & 8824 & 1.6000 & (3) \\
695.203 & 1880 & -0.3411 & (4) & 1297.715 & 8966 & 2.0352 & (3) \\
695.290 & 1881 & -0.3176 & (4) & 1299.497 & 8987 & 2.0000 & (3) \\
696.235 & 1892 & -0.2000 & (4) & 1300.608 & 9000 & 2.0705 & (3) \\
696.314 & 1893 & -0.2705 & (4) & 1309.640 & 9106 & 2.3294 & (3) \\
697.240 & 1904 & -0.3764 & (4) & 1326.679 & 9306 & 2.7882 & (3) \\
697.320 & 1905 & -0.4352 & (4) & 1327.704 & 9318 & 2.8470 & (3) \\
771.607 & 2779 & -0.4705 & (2) & 1329.663 & 9341 & 2.8941 & (3) \\
862.444 & 3848 & -0.8000 & (4) & 1333.406 & 9385 & 2.9294 & (3) \\
862.525 & 3849 & -0.8470 & (4) & 1334.691 & 9400 & 3.0470 & (3) \\
863.469 & 3860 & -0.7411 & (4) & 1335.790 & 9413 & 2.9764 & (3) \\
863.556 & 3861 & -0.7176 & (4) & 1343.382 & 9502 & 3.2941 & (3) \\
875.52 & 4002 & -0.9647 & (4) & 1361.351 & 9713 & 3.6941 & (5) \\
877.40 & 4024 & -0.8470 & (4) & & & & \\
\hline
\hline
\end{tabular}}
\end{center}
\end{table}
\medskip

\clearpage

~
\vspace{8cm}

\includegraphics{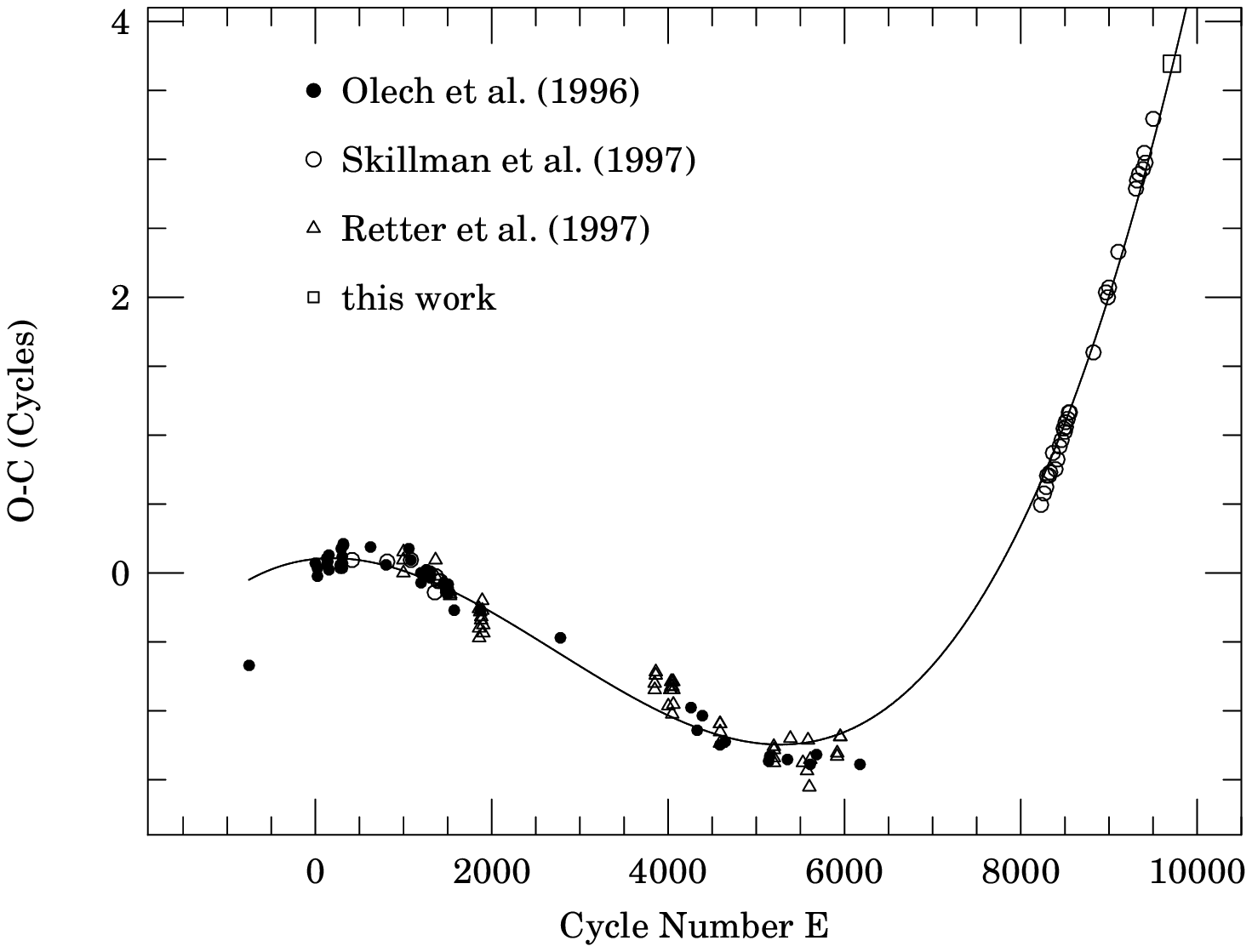}

\begin{figure}[h]
\caption{\sf $O-C$ diagram for times of maxima of the 0.085-day
modulation observed in the years 1994-1996. The $O-C$
residuals were calculated from the ephemeris (3). Black dots represents
data used by Olech et al. (1996), open circles maxima given by Skillman
et al. (1997), open triangles data presented in Retter et al. (1997)
and open squares maxima determined in this work.}
\end{figure}

To trace the 0.085-day period behavior in 1997 we prewhitened the raw
light curve of V1974 Cygni using second term of equation (1). From the
resulting light curve we derived 8 times of minima of 0.085-day
periodicity. They are shown in Table 4 together with $O-C$ residuals
computed from the following ephemeris:

\begin{equation}
{\rm HJD_{max}} = 2450634.3639 ~+~ 0.084632~E
\end{equation}
\hspace*{220pt} ${\pm}0.0068 \hspace*{10pt} {\pm}0.000080$ \\
\bigskip

\begin{table}[h]
\caption{\sc The 1997 Times of Minima of the 0.085-day       
Periodicity}
\vspace{0.1cm}
\begin{center}
\begin{tabular}{||c|r|r||c|r|r||}
\hline
\hline
HJD & E & $O-C$ & HJD & E & $O-C$ \\
2449000+ & & cycles & 2449000+ & & cycles\\
\hline
1634.366 & 0 &   0.0248 & 1641.477 & 84 &  0.0474\\
1639.431 & 60 & -0.1278 & 1642.408 & 95 &  0.0479\\
1640.456 & 72 & -0.0165 & 1642.495 & 96 &  0.0759\\
1641.389 & 83 &  0.0076 & 1645.446 & 131 & -0.0554\\
\hline
\hline
\end{tabular}
\end{center}
\end{table}

Our 1997 observing run, which covers over 100 cycles, is too short to
derive any conclusions about the 0.085-day period changes. Its mean value
obtained from the $O-C$ residuals is $0.084632(80)$ day i.e.
$121.87\pm0.12$ min which is consistent with $122.0\pm0.2$ min value
obtained from analysis of AoV power spectra.

As we mentioned earlier the 0.0813-day period was stable in the years
1994-1997 and its value was $117.0126\pm0.0003$ min. The 0.085-day
periodicity showed a more complicated behavior. It was decreasing form
the middle of 1994 until the beginning of 1995. Semeniuk et al. (1995)
and Skillman et al. (1997) reported a decrease as fast as $\dot P = -7.8
\times 10^{-7}$ but our $O-C$ diagram composed from all data available
in the literature indicates a much slower decrease with a rate of $\dot
P = -0.3 \times 10^{-7}$.

From the beginning of 1995 the 0.085-day period started to increase and it
was increasing until October 1996. Skillman et al. (1997) reported
an increase with a rate $\dot P = 5 \times 10^{-7}$ in 1996 but our $O-C$
diagram shows that the rate was not higher than $\dot P = 1 \times 10^{-7}$.

The most rapid change of the period was observed between October 1996
and July 1997. In this interval the period changed from 122.67 min
(Skillman et. al. 1997) to 121.9 min. The $\dot P$ during this period of
time could be as high as $-1.6 \times 10^{-6}$ and certainly not slower
than $-7 \times 10^{-7}$.

\vspace{13.2cm}

\includegraphics{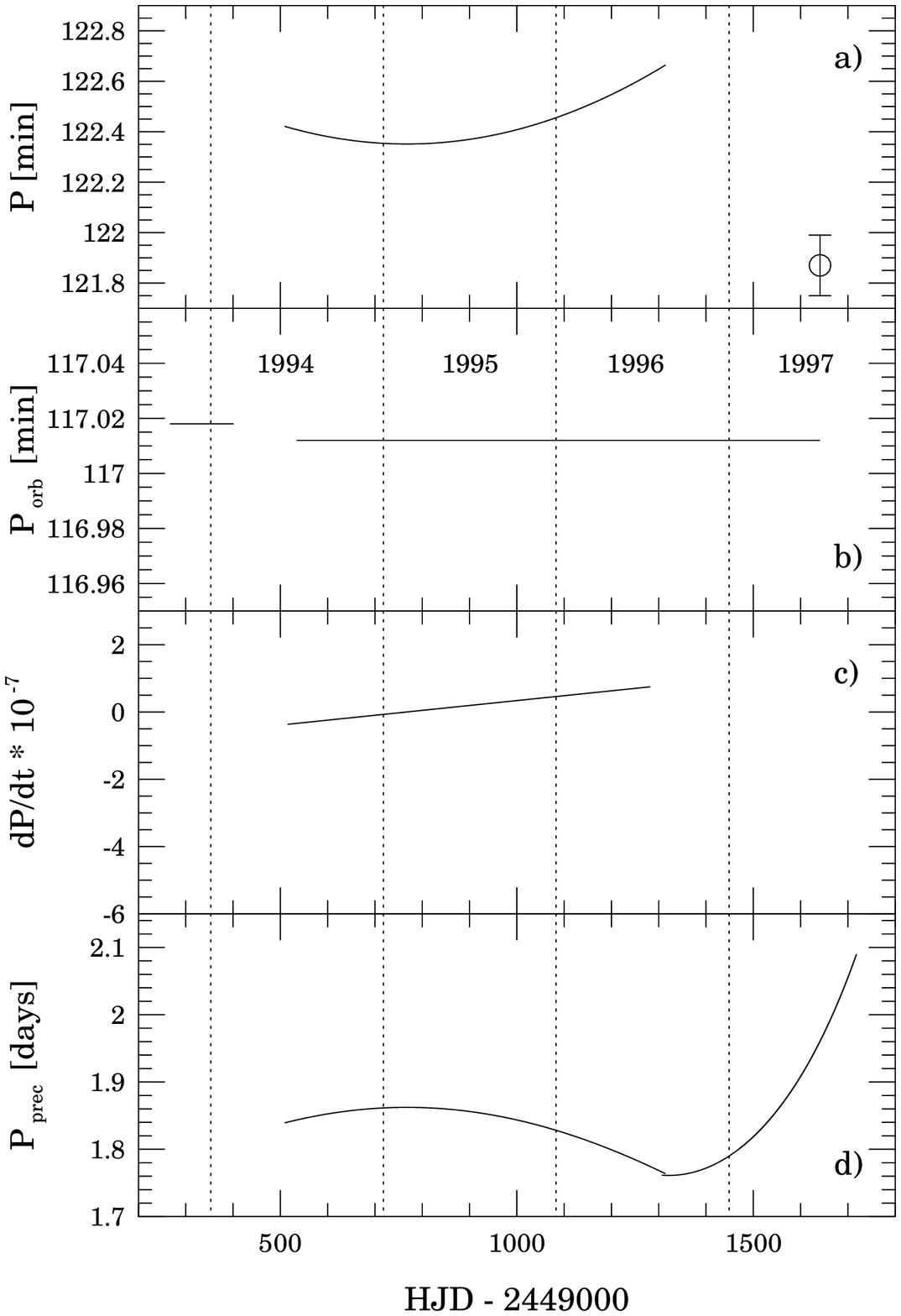}

\begin{figure}[h]
\caption{\sf Evolution of two periodicities in the light curve of V1974
Cygni. ~{\bf a)} Changes of the 0.085-day period in years
1994-1997. ~{\bf b)} Changes of the 0.0813-day period in years
1993-1997. ~{\bf c)} Changes of the time derivative of the
0.085-day period.~ {\bf d)} Changes of the beat period between 117 and
122 min periods.}
\end{figure}
\clearpage

\section{Interpretation of the signals}

Fig. 8 summarizes the results presented in this paper. It shows the
evolution of both 0.085 and 0.081-day periodicities, their changes and
evolution of the period of the disc precession assuming that V1974 Cygni
is SU UMa type variable (see discussion below).

There is consensus in the literature concerning the interpretation of
the 0.0813-day signal. Its discoverers, DeYoung and Schmidt (1994),
suggested that it can be the orbital period of the binary system
containing a white dwarf and a small main sequence star. The following
papers by Semeniuk at al. (1995), Leibowitz et al. (1995), Olech et al.
(1996), Skillman et al. (1997) and Retter et al. (1997) confirmed this
conclusion.

Our investigation also confirmed the previous results. As was shown by
Semeniuk et al. (1995) the value of the orbital period in 1993 was
0.0812623 days. In the first part of 1994 the period decreased by 0.4
seconds as was interpreted as a possible mass loss from the binary
system. We have shown that in the years 1994-1997 the 0.085-day period
was stable and its exact value was equal to $0.08125873(23)$ days
($117.0126\pm0.0003$ min) supporting the hypothesis that it is indeed
the orbital period of the binary system.

On the other hand there are two possible interpretations of the longer
0.085-day period. Semeniuk et al. (1994) who discovered it, suggested
that it can be the rotation period of the white dwarf. According to that
paper the evolution of the V1974 Cygni resembled the behavior of
intermediate polars and particularly the famous Nova V1500 Cygni. Prior
to the nova explosion, V1974 Cygni was a magnetized, synchronously
rotating AM Her type binary. During outburst the spin/orbit synchronism 
was broken and the 0.085-day period, decreasing in 1994, may be the spin
period of the white dwarf now evolving into synchronism with the orbital
cycle. According to Semeniuk et al. (1995) the predicted time scale for
equalization of both periods was $13\pm1$ yr.

Observations from 1995 analyzed by Olech et al. (1996) showed, however,
that the 0.085-day period stopped to decrease and since the middle of
July 1995 seems to be increasing. Olech et al. (1996) pointed out that
changes of the time derivative of spin periods are known in intermediate
polars. FO Aqr is a good example of such objects (Osborne and Mukai
1989, Kruszewski and Semeniuk 1993). An additional argument for the spin
nature of the 0.085-day period given by Olech et al. (1996) was a
characteristic shape of the 1995 light curve of V1974 Cygni. A distinct
feature of these light curves was a double structure of the maxima which
is often observed in polars, possibly due to the existence of two
magnetic poles (e.g. AM Her itself, Olson 1997)

Another explanation of the 0.085-day periodicity assumes that it might
be the superhump period of a SU UMa type star (Leibowitz et al. 1995,
Skillman et al. 1997, Retter et al. 1997). SU UMa stars are
non-magnetized dwarf novae showing outbursts on a time scale of days and
weeks and superoutbursts on a time scale of months. During the
superoutburst a tidal instability causes the precession of the accretion
disc. The presence of both precession and orbital periods causes the
appearance of the third, so-called superhump period which is a few
percent longer than the orbital period. In this interpretation V1974
Cygni is not a typical SU UMa star but rather a member of a newly
discovered subgroup of these variables called permanent superhumpers
(Skillman and Patterson 1993). These systems are characterized by mass
transfer rates as high as $\dot M = 6 \times 10^{16}$ g~s$^{-1}$ (Osaki
1996) and thus are understood as nova-like stars below the period gap
that are thermally stable but tidally unstable, i.e. being in permanent
superoutburst.

Assuming that V1974 Cyg is a permanent superhumper we can directly
compute its period of the disc precession which is a beat period between
orbital and superhump periods. The behavior of the precession period in
the years 1994-1997 is shown in lower panel of Fig. 8. Such rapid
changes of the accretion disc are consistent with recent models of SU
UMa systems (see for example Kornet and R\'o\.zyczka 2000 or Buat-Menart
and Hameury 2002)

About ten years ago the diversity of behavior observed in cataclysmic
variables seemed to be quite small. These variables were classified into
few distinct groups which were clearly visible in the $P_{orb}-\dot M$
diagram (see for example Fig. 3 of Osaki 1996). There was a group of SU
UMa stars below the period gap and at longer orbital periods one can
find three other groups of U Gem, Z Cam and nova-like stars, listed
according to the increasing $\dot M$.

In the mid-1990s the situation started to become more complicated. The
SU UMa group itself was divided into several subgroups as ER UMa stars
(Kato and Kunjaya 1995) and WZ Sge stars (Pych and Olech 1995, Kato et
al. 1996). Additionally Nogami et al. (1998a, 1998b) and Uemura et al.
(2001) discovered the first dwarf novae in the middle of the period gap.
The stars with short orbital periods and high accretion rates were found
to be permanent superhumper systems (Skillman and Patterson 1993). 
Superhump-like variations were also detected in ultra-short orbital
period systems containing helium secondary (e.g. AM CVn, V485 Cen, HP
Lib or  1RXS J232953.9+062814 - Skillman et al. 1999, Olech 1997,
Patterson et al. 2002, Skillman et al. 2002) and possibly a black hole
primary (e.g.  KV UMa, Uemura et al. 2002)

Permanent superhumps were also thought to be the reason for periodic light
variations in other high accretion binaries on both sides of the period gap.
At longer periods positive and negative superhumps were detected in
nova-like variables (for example V592 Cas and V751 Cyg - Taylor et al.
1998, Patterson et al. 2001).

Finally, permanent superhumps were assumed as the main cause of short
period variations of the light curves of classical nova systems lying
on both sides of the periodic gap (e.g. V603 Aql, CP Pup and V1974 Cyg
itself - Thomas 1993, Patterson and Warner 1998, Retter and Leibowitz 1998).

Retter and Leibowitz (1998) found that these three classical novae are
characterized by high accretion rates ($\dot M$ from $8\times 10^{16}$
to $8 \times 10^{17}$ g~s$^{-1}$) and thus their periodic variations
could be interpreted as permanent superhumps. The most interesting case
is V1974 Cygni itself. According to Retter and Leibowitz (1998), it
should undergo transition from a permanent superhumper to the ordinary
SU UMa star around the year 2000 (with an uncertainty of about 2 years).

To summarize, the latest results obtained for several cataclysmic
variables strongly supports the hypothesis that 122-min periodicity
observed in the light curve of V1974 Cygni is a superhump period. The
period changes and the changes of the period derivative are known in
both SU UMa stars and polars. But the temporal evolution of the 122-min
period especially in 1996-1997 was quite rapid. Such rapid changes of
the period (sometimes as high as $\dot P$ around $10^{-5}$) arising from
changes in the behavior of the accretion disc are observed in SU UMa
systems. On the other hand the $0.6-0.8$ min change of the 122-min
periodicity which occurred between 1996 and 1997 would be problematic
from the point of view of the hypothesis assuming that we observe the
rotation period of a magnetized white dwarf. Such a rapid change would
require a change in the rotational kinetic energy of the white dwarf as
large as $10^{42}$ ergs. This argument was also given by Skillman et al.
(1997) based on the period changes of V1974 Cygni in 1995-1996. In our
work, however, we showed that these changes were not as abrupt as was
previously thought, thus the change of the rotational energy was not so
large. In the end of 1996 the 122-min period started to decrease more
rapidly and thus we again should expect a change of the rotational
energy at the level of $10^{42}$ ergs which is higher than the entire
energy radiated by the star in the October 1996 - July 1997 interval.

Finally, we conclude that the explanation of the 122-min periodicity
involving the permanent superhumps phenomenon is more consistent with
observations than assumption that we are observing the rotation of
a white dwarf. However, the beginning of the new millennium is the
ideal time for checking both hypotheses. The lack of the 122-min
periodicity in the present light curve of Nova Cygni 1992 would support
the superhump explanation indicating that the object became an
ordinary SU UMa star and the presence of this period in circular
polarization would validate the connection of the period with white
dwarf rotation.

\bigskip

\noindent {\bf Acknowledgments.} We would like to thank to Prof. J\'ozef
Smak and Chris O'Connor for reading and commenting on the manuscript.
Author was supported by the Polish State Committe for Scientific
Research (KBN), grant 2P03D02422, and a Foundation for Polish Science
stipend for young scientists.

\end{document}